\documentclass[journal=jctcce,manuscript=letter]{achemso}

\SectionNumbersOn
\usepackage[utf8]{inputenc}
\usepackage[T1]{fontenc}
\usepackage{amsmath}
\usepackage{amssymb}
\usepackage{amstext}

\usepackage{siunitx}
  \DeclareSIUnit\angstrom{\textup{\AA}}

\usepackage[table]{xcolor}

\setkeys{acs}{
  maxauthors=10,
  articletitle=true,
  chaptertitle=true,
  doi=true,
}

\usepackage{enumerate}
\usepackage[version=4]{mhchem}
\usepackage{ytableau}
\usepackage{braket}
\usepackage{csquotes}
\usepackage{array}
\usepackage{booktabs}
\usepackage{cellspace}
\usepackage{graphicx}
\newcolumntype{L}{>{$}c<{$}}
\newcolumntype{R}{>{$}r<{$}}
\definecolor{shaderow}{gray}{0.92}

\author{Arta A. Safari}
  \affiliation{ETH Zürich, Laboratorium für Physikalische Chemie, 8093 Zürich, Switzerland}
  \email{arta.safari@phys.chem.ethz.ch}
\author{Markus Reiher}
  \affiliation{ETH Zürich, Laboratorium für Physikalische Chemie, 8093 Zürich, Switzerland}
  \email{markus.reiher@phys.chem.ethz.ch}

\title{Lanthanide Crystal Field Theory in the J-Mixing Regime: f$^\text{n}$ Configurations}

\begin{document}

\begin{abstract}

The low-energy sector of the lanthanide $f^n$ shell is to first approximation
    governed by the interplay of three interactions: Coulomb repulsion,
    spin-same-orbit coupling and the ligand crystal field potential.
To rationalize the energy landscape in different chemical environments, it is
    desirable to extract the phenomenological crystal field parameters from
    experimental data; however, in practice, high-dimensional fitting procedures complicate their routine
    determination.
Here, we present a computational extraction procedure based on real tensorial sets and the $U(7)$
    canonical chain that, unlike existing operator equivalent
    methods, takes all $J$ multiplets of the lowest-energy $LS$ term into
    account, ensuring meaningful results even in the presence of considerable
    $J$-mixing.
Two series of lanthanide compounds, \ce{[Ln(III)(C8H8)2]-} and
    doped Ln(III):\ce{Y2O3} (Ln $\in$ \{Ce, Pr, Nd, Sm, Eu\}) serve as examples
    for high- ($D_{8h}$) and low-symmetry ($C_2$) scenarios.
Consequences for the generalisation toward actinides are briefly discussed.

\end{abstract}

\pagebreak

\section{Introduction}
In view of the technological relevance of lanthanide compounds for efficient
    lighting technologies\cite{tessitore2023} and single-molecule
    magnets\cite{chilton2022},
    rationalizing lanthanide crystal fields in different chemical environments
    remains an important, yet challenging task.
Crystal field (CF) parameters can, in principle, be obtained by fitting
    experimental spectra against a model Hamiltonian; however, in practice, the
    large number of free parameters renders fitting procedures for low-symmetry
    complexes inherently unstable, even when employing orthogonal polynomials.\cite{judd1989}
First-principles exact-diagonalization protocols such as the complete active space self-consistent
    field (CASSCF)\cite{roos1980} combined with perturbation
    theory\cite{andersson1992,angeli2002} do not suffer from these limitations.
They are therefore key to complement experimental CF parameter
    determination.

Two established approaches to obtain CF parameters from CASSCF calculations are ab initio ligand
    field theory\cite{jung2017} and the pseudospin decomposition.\cite{ungur2017}
Whereas the former extracts CF parameters at the molecular orbital level, the latter
    employs a projection of the CASSCF wave function onto a single effective
    pseudospin manifold $^{2S+1}L_J$.
For most trivalent lanthanide ions the lowest-energy Russel-Saunders term is
    energetically well-separated from other terms with the same $LS$ and different
    $J$.
Under these conditions, CF parameters obtained with operator equivalent
    techniques such as the pseudospin decomposition are representative of experimental
    measurements in the low-energy sector.
Notable deviations from pure $J$ states arise, however, in samarium and europium
    compounds, or more generally when approaching the strong-field regime
    that is reached, for example, in lanthanide oxides or fluorides.\cite{lueken1999}
For strong-field compounds it is more accurate to project the CASSCF wave functions
    onto a model space comprising all $J$-states of the lowest-energy
    $LS$ term.

Recently,\cite{sergentu2026} the issue of $J$-mixing for the $f^1$
    configuration of the cerocene anion has been re-examined within the operator
    equivalent framework.
Following effective Hamiltonian theory for magnetic interactions,\cite{malrieu2013}
    the authors established a one-to-one correspondence between analytical and
    numerical matrices in the complete $\ket{L m_L S m_S}$ and $\ket{L S J m_J}$ bases.
In this work, we shall follow a related, but somewhat different
    effective Hamiltonian strategy to allow for a description of
    the entire lanthanide $f^n$ shell in arbitrary crystal fields.
Rather than calculating expectation values over numerical wave functions, a
    group-theoretical transformation from the many-body configuration-state
    basis to the model space is sought.\cite{safari2025}
Conceptually, this correspondence relies on a distinction between formal
    angular momentum assigned to (a linear combination of) basis functions and
    angular momentum carried by a physical state, which is measured by an expectation
    value.
As will be demonstrated below, separating the two allows strong- and weak-field
    limits to be treated on equal footing.
Another complication arises from the orientation dependence of CF parameters.
While generally there is no unique choice in the absence of point group
    symmetry, in many cases of practical interest the
    representation theory of the rotation group provides a choice that
    significantly reduces the complexity of the arising equations.

Already prior to the advent of electronic computers, the $f^n$ interaction
    matrices were formed and diagonalized using the traditional
    operator techniques of atomic
    spectroscopy.\cite{racah1942a,racah1942b,racah1943,racah1949,judd1998}
These had to capture the intrinsic complexity of the
    entire problem, necessarily leading to an
    intimidating machinery involving uncommon groups such as $G_2$, as well as
    the much-dreaded fractional parentage
    coefficients.\cite{racah1943,racah1949,harter1973}
It ought to be mentioned that despite much theoretical progress,\cite{judd1997} certain
    simplifications arising within the $f^n$ shell remain to this day not fully
    understood.
Extraction of CF parameters from the $^{2S+1}L_{|S-L|,\ldots,S+L}$ manifold
    does, however, not require all of these tools and some can be replaced by more efficient
    alternatives, leading to significant simplifications among which a few
    noteworthy are the following:

\begin{enumerate}
    \item The low-energy $f^n$ states with maximum total spin and orbital momentum,
        henceforth called \enquote{Hund-states}, are uniquely characterised by a
        single $SO(3)_L$ and $SO(3)_S$ label.
    Since the Coulomb interaction is scalar in $L$ and $S$, that is, acts as a constant
        shift within a given $LS$ term, it cancels in energy differences within the
        Hund manifold; thus, the model Hamiltonian in this space only contains
        one-electron spin-orbit and CF parameters. One may relax
            this restriction and consider all $L$ allowed for a given $S^\text{Hund}$ without introducing
            label multiplicity problems. Then, in addition to
            the aforementioned one-electron parameters, the $E3$ part of the
            Coulomb interaction can be extracted as well. It does not seem possible to
            obtain $E1$ and $E2$ without including into the model space all high-energy
            configurations with differing $S$ and re-introducing the groups $SO(7)$ and $G_2$.
    \item Difficult to implement fractional parentage coefficients are avoided by
        expressing the tensor method in the framework of $U(n)$ canonical chains,
        as originally advocated in refs. \citenum{harter1973} and \citenum{harter1976}.
    \item Easier symmetry-adaptation to the point groups and generally simpler
        ligand field matrices are obtained, when the irreducible tensor method is
        formulated in terms of real spherical harmonics, essentially hydrogenic angular functions
        with a different normalisation, instead of the usual complex harmonics.
    In light of the high coordination numbers and little to no point group symmetry
        that most molecular lanthanide complexes exhibit, the benefit of point group
        adaptation may seem negligible.
    However, a symmetry-adapted formalism remains useful
        for many solid-state systems, where symmetry reduction tends to be less
        severe.\cite{morrison1982}
    Moreover, a local idealized symmetry analysis has always been beneficial for
        the understanding of complicated electronic structures in transition metal
        complexes.
\end{enumerate}

These observations and their exploitation are elaborated upon hereinafter.
In section 2, we summarize the real tensor method originally put forth by
    Harnung and Schäffer.
Section 3 then deals with the translation to second quantization and
    $U(n)$ canonical chains, before presenting two series of lanthanide
    compounds with high and low symmetry, respectively,
    \ce{[Ln(III)(C8H8)2]-} and doped Ln(III):\ce{Y2O3} (Ln $\in$ \{Ce, Pr, Nd, Sm,
    Eu\}), in section 4.
Limitations of the present formalism in the treatment of actinides are also
    analysed using the example of \ce{NpF6}, followed by our conclusions in section 5.

\section{Real Irreducible Tensorial Sets}
\label{sec:real-sets}

The formalism of real tensorial sets in $\mathbb{R}^3$ builds upon the tensor
    method outlined by Fano and Racah in ref. \citenum{fano1959} for the chain
    $SO(3) \supset SO(2)$.
We therefore provide a brief summary of a few important definitions and facts.
The conventional irreducible tensor method for $SO(3)$ is formulated in terms of complex,
    contrastandard, irreducible sets $|\alpha j\}$ whose representations
    $\mathfrak{D}^{[j]}$ are carried by the complex, contrastandard spherical harmonics
    $\mathfrak{Y}^{[j]}_m$.
Following ref. \citenum {judd1997}, the rotation groups belonging to
    spin and orbital momentum are distinguished by subscripts $S$ and
    $L$, respectively.
In algebraic work, the homomorphism $SO(3)
        \cong SU(2)$ is mostly inconsequential and will be ignored here.
The standard set $\{ \alpha j|$ transforms contragrediently to $|\alpha j\}$ and is
    obtained from the contrastandard set by the operator $\hat{\mathrm{U}}$.
Standard representations $\mathfrak{D}^{(j)}$ are related to
    $\mathfrak{D}^{[j]}$ by complex conjugation.
Contragredience plays a central role in the tensor method, since contraction
    of contragredient pairs leads to group invariants.

The unitary matrix $\mathbf{U}^j$ is real and has the standard form for any set
    \begin{equation}
        \label{eq:umat}
        U_{mm'}^j = (-)^{j - m'} \delta(m, -m'),
    \end{equation}
    i.e., it is symmetric for integer $j$ and antisymmetric for half-integer
    $j$.
If and only if $j$ is integer, the representations can be transformed to real form; $\mathbf{U}^j$
    then becomes equal to unity.
For both integer and half-integer $j$, it is always possible to transform
    $\mathfrak{D}^{[j]}(\hat{J}_y, \pi)$, the representation of a \SI{180}{\degree} rotation around $y$,
    equal to $\mathbf{U}^j$.
Now by eq. \ref{eq:umat}, $\mathfrak{D}^{[j]}(\hat{J}_y, \pi)$ is then real and
    any transformation preserving this relationship must also be
    real.\cite{fano1959,harnung1973}
This fact will determine how we align the molecular structures with the
    Cartesian axes in section 4.

An irreducible tensorial set with standard representations of the generators $\hat{J}_x$,
    $\hat{J}_y$, $\hat{J}_z$ can be constructed from a sequence of eigenvalue
    problems.
First, the matrix belonging to $\hat{J}^2$ is diagonalized and the resulting
    eigenvectors used to transform $\mathfrak{D}^{[j]}(\hat{J}_z, \phi)$ into block-diagonal form.
Following diagonalization of $\hat{J}_z$ and assuming no degeneracies, the
    resulting tensorial sets are now irreducible, labelled by
    $jm$, with $(2j + 1)$ free phases per $j$-block to be fixed.
    Of these, $2j$ relative phases $\exp(i\Delta\gamma_m)$ can be adjusted by enforcing
    $\mathfrak{D}^{[j]}(\hat{J}_y, \theta)$ to be complex Hermitian and
    the $\hat{J}_x$ matrix real and positive.
A final global phase per $j$-block is independent of $m$ and assigned such
    that the action of the linear operator $\hat{\mathrm{U}}$ coincides with the antiunitary time-reversal operator
    $\hat{\vartheta}$, that is
    \begin{equation}
        \hat{\mathrm{U}} \ket{j m} = \exp(i\gamma_j) \hat{\vartheta} \ket{j m}
    \end{equation}
    with $\gamma_j = 0$.

In atomic spectroscopy, the labels $jm$ corresponding to $SO(3)
    \supset SO(2)$ arise naturally to reflect the spherical symmetry of atoms.
Another advantage of the complex basis is that it is defined for integer and half-integer
    representations alike.
However, many molecular problems are better described by the chain $SO(3)
    \supset P$, where $P$ is a finite point group.
As was long realized in crystal field theory,\cite{griffith1961}
    point group adaption is easier in real than in complex bases, and to this
    end, Harnung and Schäffer exploited the possibility to choose the integer representations
    as real, carried by real spherical harmonics.\cite{harnung1972b}
Another benefit of the real basis is that it often leads to simpler interaction
    Hamiltonians.
Consider for example an octahedral crystal field acting on $d$
    orbitals.
In a real basis, the familiar $t_{2g}$ / $e_g$ block-structure is immediately
    obtained, whereas in the complex basis off-diagonal mixing
    occurs.\cite{silver2013}

Incidentally, the real tensor method reflects the design of
    a scalar-relativistic CASSCF program: while the orbitals could in principle
    be complex, they are chosen to be real and paired with complex spin
    functions, ensuring real arithmetic.

When spin-orbit coupling is added, it can be advantageous to work in the
    total angular momentum $\ket{J m_J}$ basis, where this interaction is diagonal.
Again, if and only if $m_J$ is integer-valued, a real basis can be chosen,
    otherwise one has to employ a complex basis.
For simplicity, all states in the spin-orbit-coupled basis will be treated with
    complex labels.

The real spherical harmonics $\mathfrak{C}^l_{\lambda \text{\c{c}}}$
    normalised to $4\pi/(2l + 1)$ over the unit sphere
     are denoted in solid harmonic form
    \begin{equation*}
    \mathfrak{H}^l_{\lambda \text{\c{c}}} = r^l \mathfrak{C}^l_{\lambda \text{\c{c}}}.
    \end{equation*}
Since for real functions the transition to contragredience is mediated by the
    $(2l+1)$-dimensional unit matrix, no brackets or braces around the $l$ label
    are necessary.
The solid harmonics are eigenfunctions of $\hat{L}^2$ and $\hat{L}^2_z$, with
    eigenvalues $l(l+1)$ and $\lambda^2$, respectively, and integer $0 \leq
    \lambda \leq l$.
They are, however, not eigenfunctions of $\hat{L}_z$.
With a different normalization, the solid harmonics occur for instance in the Schrödinger equation for hydrogen-like atoms.
Explicit algebraic expressions\cite{harnung1972a,harnung1972b} in a right-handed spherical coordinate
    system ($r$, $\theta$, $\phi$), show that they are proportional to either
       $r^l \sin^\lambda(\theta)\sin(\lambda\phi)$ or $r^l
       \sin^\lambda(\theta)\cos(\lambda\phi)$.
Fortunately, all transformation properties relevant for
    integration with the Wigner-Eckart theorem can be encoded in the labels $l,
    \lambda, \text{\c{c}}$, with $\text{\c{c}} = s$ for sine- and $\text{\c{c}}
    = c$ for cosine-type functions.
Explicit expansions are hence rarely needed in practice.
Solid harmonics with $\lambda = 0$, the so-called zonal harmonics, have
    rotational symmetry around the $z$-axis.
They are usually grouped with the cosine-type functions $\text{\c{c} = c}$, but carry the
    special subscript $\sigma$.

The real basis for SO(3) is denoted by letters
    $\ket{l t} \equiv \ket{l \lambda \text{\c{c}}}$ with standard order
    \begin{equation*}
        \ket{l\sigma}, \ket{1, 1s}, \ket{l,1c}, \ket{l, 2s}, \ldots, \ket{l, \lambda c}
    \end{equation*}
    and standard order is assumed in all presented equations and
    tables.
Complex and real representations are related by a constant unitary matrix,
    which in the Fano-Racah phase convention outlined above takes the form
\begin{equation}
    \begin{pmatrix} \ket{l, \sigma} & \ket{l, \lambda s} & \ket{l, \lambda c} \end{pmatrix} =
    \begin{pmatrix} \ket{j, \lambda} & \ket{j, 0} & \ket{j, -\lambda} \end{pmatrix}
    \begin{pmatrix}
        0        & (-)^\lambda \dfrac{(-i)^{j+1}}{\sqrt{2}} & (-)^\lambda \dfrac{(-i)^{j}}{\sqrt{2}} \\
        (-i)^{j} & 0                                        & 0                                      \\
        0        & -\dfrac{(-i)^{j+1}}{\sqrt{2}}              & \dfrac{(-i)^{j}}{\sqrt{2}}               \\
    \end{pmatrix}.
\end{equation}
Note the tensorial sets are arranged in rows instead of columns.

Introducing real basis functions for the three-dimensional rotation group
    leads to a new set of coupling coefficients, which were worked out in
    detail in ref. \citenum{harnung1972b}.
The conventional vector coupling coefficients for the basis $SO(3) \supset
    SO(2)$, which reduce the direct product of two sets
    \begin{equation}
        \label{eq:cg-coeffs}
        \left\{ |j_1\} \otimes |j_2\} \right\}^{ [ j_3 ] }_{m_3} = \sum_{m_1 m_2} \ket{j_1 m_1; j_2 m_2}\Braket{j_1 m_1; j_2 m_2 | j_3 m_3},
    \end{equation}
    differ by more than a phase under certain permutations and time reversal of angular momenta
    and are therefore re-expressed in terms of the more symmetric Wigner 3-$jm$ symbol
    \begin{equation}
        \Braket{j_1 m_1; j_2 m_2 | j_3 m_3} = (-)^{2j_2 + j_3 - m_3} \sqrt{2j_3 + 1}
        (-)^{j_1 + j_2 + j_3} \begin{pmatrix} j_1 & j_2 & j_3 \\ m_1 & m_2 & -m_3 \end{pmatrix}.
    \end{equation}
Analogously, in the real basis the coupling coefficients are defined through
    what Harnung and Schäffer call the 3-$l$ symbol
    \begin{equation}
        \Braket{l_1 t_1; l_2 t_2 | l_3 t_3} =  \sqrt{2l_3 + 1}
        (-)^{l_1 + l_2 + l_3} \begin{pmatrix} l_1 & l_2 & l_3 \\ t_1 & t_2 & t_3 \end{pmatrix}
    \end{equation}
    with the same symmetry properties as the 3-$jm$ symbol.
Since the coupling coefficients are basis-dependent, we prefer and use the name 3-$lt$ symbol instead.

In case $\lambda_1, \lambda_2, \lambda_3$ are all larger than zero,
    the 3-$lt$ symbols may be computed from the standard 3-$jm$ symbols
    after rearranging the columns to fulfil $\lambda_1 + \lambda_2 = \lambda_3$,\cite{harnung1972b}
    \begin{equation}
        \label{eq:3lt}
        \begin{pmatrix} l_1 & l_2 & l_3 \\ \lambda_1 \text{\c{c}}_1 & \lambda_2 \text{\c{c}}_2 & \lambda_3 \text{\c{c}}_3 \end{pmatrix} =
            \text{phase} \cdot \dfrac{(-)^{l_3 - \lambda_3}}{\sqrt{2}}\begin{pmatrix} l_1 & l_2 & l_3 \\ \lambda_1 & \lambda_2 & -\lambda_3 \end{pmatrix}.
    \end{equation}
The phase depends on $(l_1 + l_2 - l_3) \text{ mod } 4$ and is listed
    in table \ref{tab:3lt-phases}.
\begin{table}[!ht]
    \caption{Phases required for the computation of the 3-$lt$ symbol in eq. \ref{eq:3lt}.}
    \centering
    \renewcommand{\arraystretch}{1.1}
    \begin{tabular}{ccc c cccc}
                 &                  &                  &       & \multicolumn{4}{c}{$l_1 + l_2 - l_3 \text{ mod } 4$} \\
        $\text{\c{c}}_1$ & $\text{\c{c}}_2$ & $\text{\c{c}}_3$ & \quad & $\equiv 0$ & $\equiv 2$ & $\equiv 1$ & $\equiv 3$ \\
        \hline
        $c$ & $c$ & $c$ & & $[ + ]$ & $[ - ]$                     \\
        $s$ & $s$ & $c$ & & $[ - ]$ & $[ + ]$                     \\
        $s$ & $c$ & $s$ & & $[ + ]$ & $[ - ]$                     \\
        $c$ & $s$ & $s$ & & $[ + ]$ & $[ - ]$                     \\
        $s$ & $s$ & $s$ & &         &         & $[ - ]$ & $[ + ]$ \\
        $c$ & $c$ & $s$ & &         &         & $[ + ]$ & $[ - ]$ \\
        $c$ & $s$ & $c$ & &         &         & $[ - ]$ & $[ + ]$ \\
        $s$ & $c$ & $c$ & &         &         & $[ - ]$ & $[ + ]$
    \end{tabular}
    \label{tab:3lt-phases}
\end{table}
If $\lambda_1$ or $\lambda_2$ is zero, the pre-factor $2^{-1/2}$ needs to be removed.
Another special case occurs when all three entries are zonal functions.
The corresponding 3-$lt$ symbol is only nonvanishing for even $L = l_1 + l_2 + l_3$
    and takes the value
    \begin{equation}
        \begin{pmatrix} l_1 & l_2 & l_3 \\ \sigma & \sigma & \sigma \end{pmatrix}
            = \dfrac{(L/2)!}{(L/2 - l_1)! (L/2 - l_2)! (L/2 - l_3)!} \sqrt{\dfrac{(L - 2l_1)! (L - 2l_2)! (L - 2l_3)!}{(L + 1)!}}.
    \end{equation}

In conventional angular momentum theory, contractions of two, four, six, ...
    3-$jm$ symbols define SO(3) invariants, the $3j$, $6j$, $9j$, ... symbols,
    commonly interpreted as part of the transformations between different coupling schemes.
As their names suggest, these symbols only depend on $SO(3)$ labels and remain
    identical in the real basis.
With these definitions, the Wigner-Eckart theorem for the complex basis\cite{butler1981}
    \begin{equation}
        \braket{\alpha_1 j_1 m_1|\hat{\mathfrak{O}}^{[j_2]}_{m_2}|\alpha_3 j_3 m_3} = (-)^{j_1 - m_1}
        \begin{pmatrix} j_1 & j_2 & j_3 \\ -m_1 & m_2 & m_3 \end{pmatrix}
        \braket{\alpha_1 j_1 ||\hat{\mathfrak{O}}^{[j_2]}||\alpha_3 j_3}
    \end{equation}
    assumes the real form
    \begin{equation}
        \braket{\alpha_1 l_1 t_1|\hat{\mathfrak{O}}^{l_2}_{t_2}|\alpha_3 l_3 t_3} =
        \begin{pmatrix} l_1 & l_2 & l_3 \\ t_1 & t_2 & t_3 \end{pmatrix}
        \braket{\alpha_1 l_1 ||\hat{\mathfrak{O}}^{l_2}||\alpha_3 l_3}.
    \end{equation}
The reduced matrix elements are identical, provided both operators span the
    same space.\cite{harnung1972b}

When real spherical harmonics act as operators, notably in expansions of the
    crystal field potential, here for a single electron,
    \begin{equation}
        \begin{split}
            \Braket{\alpha_1 l_1 t_1 |\hat{v}_\text{crys}|\alpha_3 l_3 t_3}
                &= \Braket{\alpha_1 l_1 t_1|\sum_{l_2}^{\{2, 4, 6\}} \sum_{t_2} \mathfrak{A}^{l_2}_{t_2} \hat{\mathfrak{H}}^{l_2}_{t_2}|\alpha_3 l_3 t_3} \\
        &= \sum_{l_2}^{\{2, 4, 6\}} \sum_{t_2} \underbrace{\mathfrak{A}^{l_2}_{t_2}
            \Braket{ \hat{r}^{l_2} } \Braket{\alpha_1 l_1||\hat{\mathfrak{C}}^{l_2}||\alpha_3 l_3 }}_{E^{l_2}_{t_2}} \begin{pmatrix} l_1 & l_2 & l_3 \\ t_1 & t_2 & t_3 \end{pmatrix},
        \end{split}
    \end{equation}
    with the radial integral $\Braket{\hat{r}^l}$, the reduced matrix element
    can be evaluated in terms of the 3-$lt$ symbol
    \begin{equation}
        \label{eq:red-elem-cart-tensor}
        \Braket{ l_1||\hat{\mathfrak{C}}^{l_2}||l_3 } = \sqrt{(2l_1 + 1)(2l_3 + 1)} \begin{pmatrix} l_1 & l_2 & l_3 \\ \sigma & \sigma & \sigma \end{pmatrix}.
    \end{equation}

\section{Group Classification in Second Quantisation}
\label{sec:uga-classification}
To combine the real tensorial formalism with complete active space methods, the
    former has to be translated to second quantization.\cite{harter1973,harter1976}
In this work, we limit ourselves to the discussion of a single atomic
    $f$-shell with $l_i = 3$.

Starting point for the group-theoretical classification\cite{judd1997} is the
    observation that the particle-number-conserving spin-orbital excitations
    $\hat{a}^{\dagger}_P \hat{a}_Q$ close under commutation
    \begin{equation}
        [\hat{a}^{\dagger}_P \hat{a}_Q, \hat{a}^{\dagger}_R \hat{a}_S]
        = \delta_{QR} \hat{a}^{\dagger}_P \hat{a}_S - \delta_{PS} \hat{a}^{\dagger}_R \hat{a}_Q
    \end{equation}
    and constitute the generators of the Lie group $U(4l + 2) = U(14)$.
Subgroups are identified by grouping these operators into subsets
    closed under commutation.\cite{judd1997}
Recognizing that the one-electron creation and annihilation operators form
    double-tensorial sets of rank $s_i = 1/2$ in spin- and rank $l_i = 3$ in orbital
    space, one can sum over the spin projection and orbital labels
    \begin{equation}
        \hat{E}_{p_1, p_2} = \sum_{m \in \alpha,\beta} \hat{a}^{\dagger}_{p_1,m} \hat{a}_{p_2,m}
        \qquad
        \hat{T}_{m_1, m_2} = \sum_{p \in \sigma,1s,1c,\ldots} \hat{a}^{\dagger}_{p, m_1} \hat{a}_{p, m_2}
        \label{eq:u14-subduction}
    \end{equation}
    to achieve the subduction $U(14) \supset U(7) \times U(2)$.
Consequently, in eq. \ref{eq:u14-subduction} the $U(14)$ labels $P,Q$ are
    replaced by $U(7) \times U(2)$ labels $\{ p, m_1 \}$, $\{ q, m_2 \}$ to arrive
    at the usual notation of \enquote{spin-free} electronic
    structure.\cite{helgaker2000}
Many molecular electronic structure packages employ this scheme due to its
    computational efficiency; for the majority of the periodic table, real-valued,
    sometimes scalar-relativistically corrected wave functions represent accurate
    approximations to the complex-valued solutions of two- or four-component
    theories.

In the traditional treatment by Racah,\cite{racah1949} the spatial orbital group $U(7)$ is further
    decomposed along the chain $U(7) \supset SO(7) \supset G_2 \supset SO(3)_L
    \supset SO(2)_L$, where $G_2$ is one of Cartan's exceptional Lie groups.
This construction has the benefit of leading to $SO(3)_L \times SO(3)_S$, the
    equivalent of $LS$ coupling.
However; as has been long
    realised,\cite{judd1998}
    for the configurations $f^5, f^6, f^7$ and their particle-hole
    counterparts, no chain ending in these groups can uniquely distinguish
    all atomic terms.
A formal solution to the labelling problem for arbitrary systems is provided by the canonical
    \enquote{Gel'fand-Tsetlin} chain\cite{gelfand1950}
    \begin{equation*}
        U(7) \supset U(6) \supset \ldots \supset U(1)
    \end{equation*}
    familiar in electronic structure as the unitary group
    approach.\cite{moshinsky1963,paldus1976a,pauncz2018}
The $N$-electron orbital functions carrying a given two-column irreducible representation of $U(n)$ encode a
    complicated product of uncoupled, real solid harmonics, whose combinatorial
    structure may be \enquote{visualized} by exploiting the one-to-one
    correspondence between $U(n)$ weights and Young tableaux.\cite{judd1998,pauncz2018}
Just as for the spherical harmonics, explicit expressions are in principle
    available, albeit unnecessarily verbose.
The labels of the canonical configuration state functions, which
    reduce to the $\{ 0,u,d,2 \}$ notation for electronic
    two-column representations,\cite{paldus1976b} again suffice for their manipulation.
At the same time, the uncoupled $SO(3)_L$ structure of the canonical basis implies
    additional steps to adapt them to total orbital momentum or physical
    symmetries other than total spin.

We note that proposals to employ the unitary group approach in crystal field problems can be traced
    back to the early 1980s,\cite{zhenyi1983} but the manual adaptation to
    additional symmetries and, at the time, underdeveloped theory to deal with spin-dependent
    multi-shell interactions are probably the reason why they did not achieve
    widespread adoption.
In a classic book,\cite{wybourne1974} canonical chains are only mentioned parenthetically
    in the section on \enquote{Simple Isoscalar Factors}: \enquote{[...] these
    groups are seldom of physical significance, though it can sometimes be useful
    to calculate first in the Gel'fand-Tsetlin basis and then make a transformation
    to a physical basis.}

The most straightforward way to adapt the canonical basis to additional
    symmetries is to diagonalize the Casimir operators of the pertinent
    Lie algebras, followed by enforcement of the desired phase convention.
Another, in principle more elegant path starts from the highest-weight state in
    the Cartan-Weyl sense, which is trivial to identify when working with complex functions
    and employs ladder operators in conjunction with orthogonalization.\cite{schlesinger1982}
All free phases are consistently fixed this way, but the action of ladder
    operators on the real basis is not simple, making the procedure unsuitable for the
    present context.

To represent the action of $\hat{L}^2$ or other tensor operators on the canonical
    basis, we shall express them in terms of $U(n)$ generators.\cite{harter1973,harter1976}
The arising integrals can then readily be evaluated using group-theoretical\cite{paldus2020a} or
    determinantal\cite{galvan2026} techniques.
Consider the rank $k$, unit tensor operator $\hat{u}^k$ acting on electron $i$;
    inserting the identity over orbital space twice, one obtains
    \begin{equation}
        \hat{u}^k (i) = \sum_{p,q}^{n_\text{orb}} \ket{p} \Braket{p | \hat{u}^k (i) | q} \bra{q}
            = \sum_{p,q}^{n_\text{orb}} \Braket{p | \hat{u}^k (i) | q} \hat{e}_{pq}.
        \label{eq:unit-tensor-expansion}
    \end{equation}
Further simplification through the Wigner-Eckart theorem
    requires the numerical orbitals encoded by $\ket{p},\ket{q}$ to carry $SO(3)_L$ labels, i.e., be in some
    sense equivalent to the solid harmonics in standard order.
In a molecular CASSCF calculation, this condition will never be met exactly;
    nevertheless, it can be fulfilled in a least-square sense by rotating the
    orbital angular momentum matrices in the molecular orbital basis as close as
    possible to their formal atomic counterparts
    \begin{equation*}
        \underset{\mathbf{O} \in SO(7)}{\text{minimize}} \qquad \dfrac{1}{3} \sum_{t \in \sigma,1s,1c}
            \big|\big| \mathbf{O}^T \mathbf{L}^{CAS}_t \mathbf{O} -
            \mathbf{L}^{atom}_t \big| \big|_2^2 .
    \end{equation*}
Now, the matrix element in eq. \ref{eq:unit-tensor-expansion} simplifies to
    \begin{equation}
        \braket{ 3 t_1 | \hat{u}^k_{t_2} | 3 t_{3} } = \begin{pmatrix} 3 & k & 3 \\ t_1 & t_2 & t_3 \end{pmatrix}
    \end{equation}
    and summing over all electrons, one arrives at linear combinations of $U(7)$
    generators
    \begin{equation}
        \sum_i^{n_\text{elec}} \hat{u}^k_{t_2} (i) = \hat{U}^k_{t_2}
        = \sum_{t_1 t_3} \begin{pmatrix} 3 & k & 3 \\ t_1 & t_2 & t_3
        \end{pmatrix} \hat{E}_{t_1,t_3}.
    \end{equation}
Note that $\hat{U}^k_{t_2}$ here is a multi-electron unit tensor operator, not the
    operator $\hat{\mathrm{U}}$ that transform a tensorial set to its
    contragredient counterpart.
Other one-electron operators are proportional to $\hat{U}^k_{t_2}$ by their reduced matrix
    element, e.g., for angular momentum operators, the proportionality factor is
    well-known to be\cite{fano1959,harnung1972b}
    \begin{equation}
        \Braket{j||\hat{j}^1||j} = i \sqrt{(2j + 1)j(j + 1)}
    \end{equation}
    and the solid harmonics acting as operators were discussed back in section
    \ref{sec:real-sets}, eq. \ref{eq:red-elem-cart-tensor}.

Second-quantized expressions for spin-dependent operators, such as spin-orbit
    coupling, are derived in a similar fashion.
In the basis $U(14) \supset U(7) \times U(2)$, the
    spin-(same)-orbit interaction is easier treated as an uncoupled rank one double tensor operator in both spin and
    orbital momentum (note the mixed use of real and complex components for
    $\hat{l}$ and $\hat{s}$, respectively)
\begin{equation}
    \begin{split}
        \hat{h}_\text{soc} &= \xi \left( \hat{l}^1_\sigma \hat{s}^1_\sigma + \hat{l}^1_{1s} \hat{s}^1_{1s} + \hat{l}^1_{1c} \hat{s}^1_{1c} \right) \\
                           &= \xi \left( -i \hat{l}^1_\sigma \hat{s}^1_0 + \hat{l}^1_{1s} \dfrac{1}{\sqrt{2}} \left( \hat{s}^1_{+1} + \hat{s}^1_{-1} \right) + \hat{l}^1_{1c} \dfrac{i}{\sqrt{2}} \left( \hat{s}^1_{+1} - \hat{s}^1_{-1} \right) \right) \\
                           &= \xi \left( \sum_{(t_2, m_2)}^{\substack{(\sigma, 0), \\(1s, \pm 1), (1c, \pm 1)}} c_{t_2, m_2} \hat{l}^1_{t_2} \hat{s}^1_{m_2}  \right).
    \end{split}
\end{equation}
As before, the identity is inserted, here over both spin and orbital spaces,
    followed by application of the Wigner-Eckart theorem,
\begin{equation}
    \begin{split}
        \sum_{t_1,t_3}^{\sigma,\ldots,3c} &\sum_{m_1,m_3}^{\pm 1/2}
        \Ket{^{3, 1/2}_{t_1,m_1}}\Bra{^{3, 1/2}_{t_1,m_1}} \hat{h}_\text{soc} \Ket{^{3, 1/2}_{t_3,m_3}}\Bra{^{3, 1/2}_{t_3,m_3}} \\
        &= \sum_{t_1,t_3}^{\sigma,\ldots,3c} \sum_{m_1,m_3}^{\pm 1/2} \sum_{(t_2, m_2)}^{\substack{(\sigma, 0), \\(1s, \pm 1), (1c, \pm 1)}} c_{t_2, m_2}
            \begin{pmatrix} 3 & 1 & 3 \\ t_1 & t_2 & t_3 \end{pmatrix} \Braket{ 3||\hat{l}^{1}||3 } \\
            &\qquad (-)^{s_1 - m_1} \begin{pmatrix} 1/2 & 1 & 1/2 \\ -m_1 & m_2 & m_3 \end{pmatrix} \Braket{1/2||\hat{s}^{1}||1/2} \Ket{^{3, 1/2}_{t_1,m_1}}\Bra{^{3, 1/2}_{t_3,m_3}}
    \end{split}
\end{equation}
    and summation over all electrons, $\sum_i \hat{h}_\text{soc}(i) = \hat{H}_\text{soc}$.
The resulting matrix elements can again be evaluated using the appropriate
    unitary group\cite{paldus2020a} or determinant formalism\cite{harter1976,galvan2026}.

After all, the model Hamiltonian in the Hund manifold using first-quantized tensorial
    notation
\begin{equation}
    \hat{H}_\mathrm{mod} = \sum_{i = 1}^{n_\mathrm{elec}} \xi(r_i)
        \sqrt{3} \left\{ \hat{l}^1(i) \otimes \hat{s}^1(i) \right\}^0_\sigma
        + \sum_{i = 1}^{n_\mathrm{elec}} \sum_{l_2}^{\{2,4,6\}} \sum_{t_2} \mathfrak{A}^{l_2}_{t_2} \hat{\mathfrak{H}}^{l_2}_{t_2}(i)
\end{equation}
    takes the second-quantized form
\begin{equation}
    \begin{split}
    \hat{H}_\mathrm{mod} &= \sum_{t_1,t_3}^{\sigma,\ldots,3c} \sum_{m_1,m_3}
    \Bra{^{3,1/2}_{t_1,m_1}} \hat{H}_\text{soc} \Ket{^{3,1/2}_{t_3,m_3}} \hat{a}^\dagger_{t_1,m_1} \hat{a}_{t_3,m_3}\\
                         &+ \sum_{l_2}^{\{2,4,6\}} \sum_{t_2}^{\sigma,\ldots,l_2 c} \mathfrak{A}^{l_2}_{t_2} \sum_{t_1 t_3}^{\sigma,\ldots,3c} \Braket{3 t_1|\hat{\mathfrak{H}}^{l_2}_{t_2}|3 t_3} \hat{E}_{t_1 t_3} \end{split}.
\end{equation}

\section{Results and Discussion}
\label{sec:results}
\subsection{Computational Details}
All calculations were performed with \texttt{OpenMolcas} 26.06.\cite{limanni2023b}
The ANO-RCC basis of VTZP quality on the lanthanide and actinide centres was
    contracted to Ln: (25s22p15d11f4g2h) / [8s7p4d3f2g1h] and
    Np: (26s23p17d13f5g3h) / [9s8p6d4f2g1h].
On the ligands, we employed F: (14s9p4d3f) / [4s3p2d1f], C/O: (14s9p4d) / [3s2p1d],
    H: (8s4p) / [2s1p] of VDZP quality and Y: (21s18p13d) / [5s4p2d] of MB quality,
    in conjunction with the Douglas-Kroll-Hess
    Hamiltonian\cite{wolf2002,reiher2006} of second order for the lanthanide systems and
    of fourth order for the actinide system.
Point charges and the \ce{Y2O3} embedded cluster model were obtained from
    the \texttt{chargedel} package\cite{bogdanov2021}, based on an experimental
    structure at room temperature.\cite{kevorkov1995}
As explained in section \ref{sec:real-sets}, the $y$-direction was chosen to
    coincide with (one of) the $C_2$ axes, because then all model and effective
    Hamiltonian matrices are guaranteed to be
    real in the employed phase convention.
State-averaged seven-orbital CASSCF calculations were performed over the lowest $(2L + 1)$
    states expected from the lowest-energy $LS$ term.
Since the model Hamiltonians employed in this work assume a common orbital basis for
    the CASSCF states, all calculations were performed without point group symmetry.
Spurious symmetry breaking was excluded by verifying that the spin-free
    energy splittings match the ones expected from the subduction $SO(3)_L
    \supset P$.
Sample input files are provided in the supporting information.

Numerical effective Hamiltonians were constructed following the standard
    procedure as outlined for instance in ref. \citenum{degraaf2016}.
The states $\ket{\Psi^\text{CAS}_i}$ relaxed in the full CAS space, which includes all
    allowed $L$ for a given $S^\text{Hund}$, were projected onto the
    $L^\text{Hund}S^\text{Hund}$ model space, afterwards orthonormalised to
    obtain $\ket{\widetilde{\Psi}^\text{CAS}_i}$ and used to express $H_\text{eff}$
    by spectral decomposition
    \begin{equation}
        H_\text{eff} = \sum_{i}^{n_\text{states}} \ket{\widetilde{\Psi}^\text{CAS}_i}
        E^\text{CAS}_i \bra{\widetilde{\Psi}^\text{CAS}_i}.
    \end{equation}
Ligand field and spin-orbit parameters were extracted in post-processing
    relying on the \texttt{numpy}\cite{harris2020}, \texttt{sympy}\cite{meurer2017},
    \texttt{scipy}\cite{virtanen2020}, \texttt{h5py}\cite{h5py}
    and \texttt{py3nj}\cite{fujiipy3nj} libraries.
Symbolic model and numerical effective Hamiltonians were first computed in $SO(3)_S
    \times SO(3)_L$ and then transformed through the standard vector coupling
    coefficients, eq. \ref{eq:cg-coeffs}, into $SO(3)_S \times SO(3)_L \supset
    SO(3)_J$ to allow for a comparison of the parameters extracted from both
    bases.
The spin-free symbolic matrices without the spin-orbit terms are small enough to be
    printed on paper and provided in PDF format.

\subsection{\ce{[Ln(III)(C8H8)2]-}}
\begin{figure}
    \centering
    \includegraphics[width=0.6\textwidth]{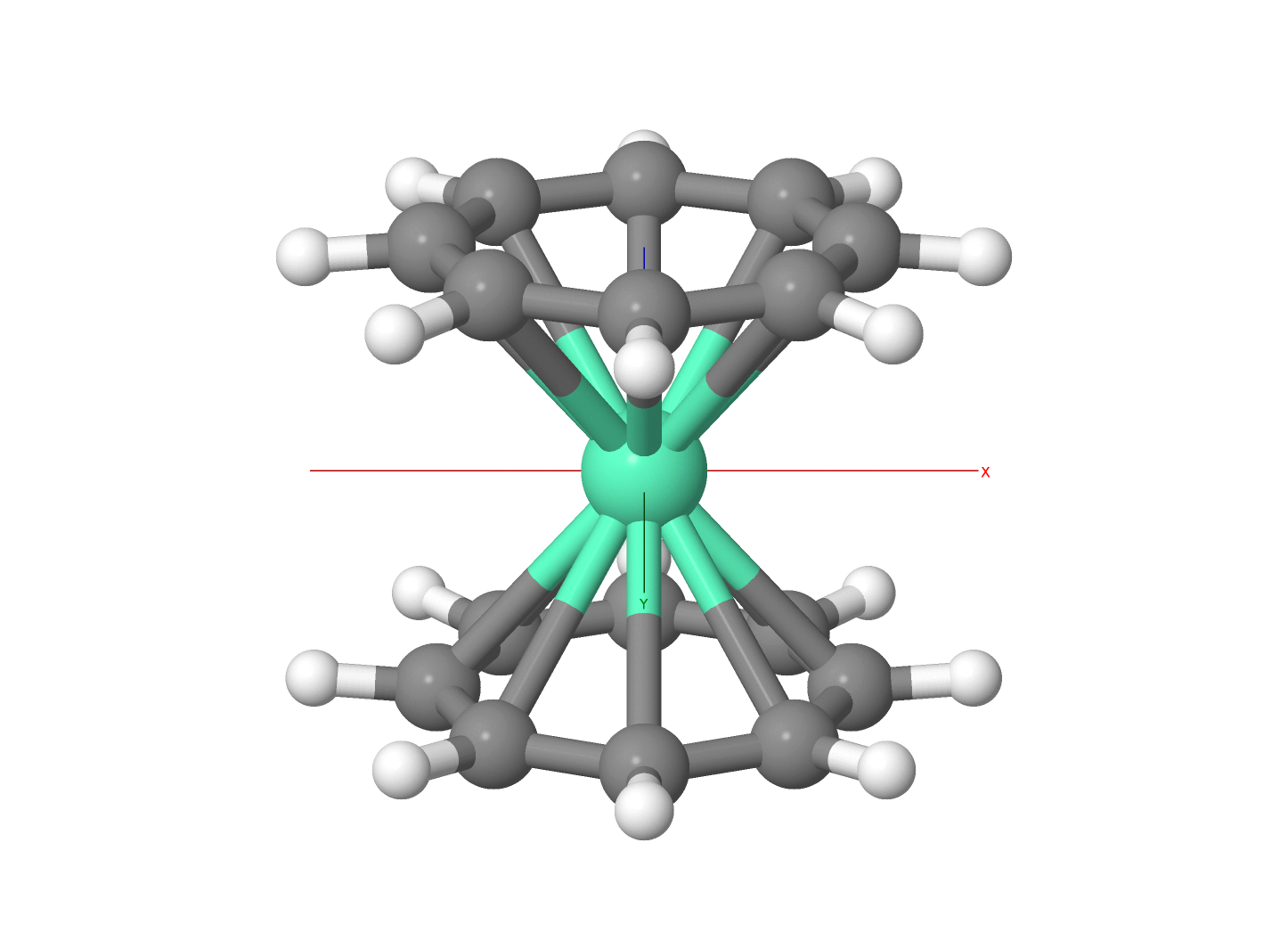}
    \caption{\ce{[Ln(III)(C8H8)2]-} sandwich complex as visualized with
        Jmol.\cite{hanson2010} \ce{Ln(III)} shown in mint turquoise, carbon in
        gray and hydrogen in white.}
    \label{fig:sandwich}
\end{figure}

As a first test case for the extraction procedure and an example of
    high-symmetry crystal fields, we consider a series of \ce{[Ln(III)(C8H8)2]-}
    sandwich complexes with $D_{8h}$ point group symmetry, see figure
    \ref{fig:sandwich}, based on ref. \citenum{sergentu2026}.
Under the chain $SO(3)_L \supset D_{8h}$, the low-energy atomic terms split as
    $F \rightarrow A_{2u} + E_{1u} + E_{2u} + E_{3u}$ (\ce{Ce^{3+}} / \ce{Eu^{3+}}) ,
    $H \rightarrow A_{2u} + B_{1u} + B_{2u} + E_{1u} + E_{2u} + 2E_{3u}$ (\ce{Pr^{3+}}, \ce{Sm^{3+}}) and
    $I \rightarrow A_{1g} + B_{1g} + B_{2g} + E_{1g} + 2E_{2g} + 2E_{3g}$ (\ce{Nd^{3+}}).
Totally symmetric in $D_{8h}$ are only the zonal harmonics
    $\mathfrak{H}^2_\sigma$, $\mathfrak{H}^4_\sigma$, $\mathfrak{H}^6_\sigma$,
    leading to a total of four free parameters including spin-orbit-coupling.
From an element-wise least-square fit, the parameters listed in table
    \ref{tab:cene-data} are obtained.

\begin{table}[!ht]
    \caption{Crystal field- and spin-orbit-parameters in cm$^{-1}$ obtained for \ce{[Ln(III)(C8H8)2]-} from
    least-square fits in the spin-free $\ket{S, Lt}$ and total angular momentum
    basis $\ket{J m_J}$.}
    \centering
    \renewcommand{\arraystretch}{1.1}
    \resizebox{\textwidth}{!}{
    \begin{tabular}{L RR|RR|RR|RR|RR}
                   & \multicolumn{2}{c }{\ce{Ce^{3+}}} & \multicolumn{2}{c }{\ce{Pr^{3+}}} & \multicolumn{2}{c }{\ce{Nd^{3+}}} & \multicolumn{2}{c }{\ce{Sm^{3+}}} & \multicolumn{2}{c}{\ce{Eu^{3+}}} \\
                   & \ket{S, Lt} & \ket{J m_J} & \ket{S, Lt} & \ket{J m_J} & \ket{S, Lt} & \ket{J m_J} & \ket{S, Lt} & \ket{J m_J} & \ket{S, Lt} & \ket{J m_J} \\
                   \cmidrule{2-11}
\rowcolor{shaderow}
        E^2_\sigma & -742   & -742   & -1189 & -1191 & -1088 & -1091 & -937  & -931  & -993  & -992  \\
        E^4_\sigma & -5706  & -5710  & -4722 & -4711 & -4169 & -4174 & -3336 & -3340 & -2755 & -2754 \\
\rowcolor{shaderow}
        E^6_\sigma & 924    & 921    & 1005  & 1004  & 862   & 862   & 589   & 589   & 600   & 600   \\
        \xi        & -      & 671    & -       & 801   & -       & 948   & -       & 1247  & -       & 1421  \\
\rowcolor{shaderow}
        b          & 0        & -4       & -25     & 0       & 8       & -1      & 2       & 10      & 0       & 0       \\
        R^2        & >0.9999  & >0.9999  & >0.9999 & >0.9999 & >0.9999 & >0.9999 & >0.9999 & >0.9999 & >0.9999 & >0.9999 \\
    \end{tabular}
    \label{tab:cene-data}
    }
\end{table}
Overall, the CF parameters obtained from the $\ket{S, Lt}$ and $\ket{J m_J}$ bases are
    almost identical, as they ought to be, since the crystal field and
    spin-orbit interactions act as separate perturbations on the free-ion
    Hamiltonian.
The remaining deviations arise from spurious orbital hybridization,
    which introduces contributions from outside the model space, as well as a small
    renormalisation error incurred by ignoring the effect of the non-Hund
    states in the model Hamiltonian.
Among these results, the spin-free ones for \ce{Pr^{3+}} are notable for
    including the largest barycenter correction $b$ in the $\ket{S,
    Lt}$ basis.
Upon examination of the spin-free wave functions, we find
    weak $L$-mixing in roots seven, eight and nine, with $3.2\%$, $1.23\%$
    and $1.23\%$ contamination of $^3H$ by the $^3F$ term.
Fits in the $\ket{J m_J}$ basis seem to absorb these deviations into other
    parameters, providing a case in point to always compare the results
    obtained from both bases.

Accounting for the conversion factors between Steven's and Wybourne's
    normalization,\cite{duros2024} allows comparison of the \ce{Ce^{3+}} data
    with the results from ref. \citenum{sergentu2026}.
Expressed in the latter convention used here, the authors reported
    $E^2_\sigma = \SI{-728.4}{\per\centi\metre}$, $E^4_\sigma = \SI{-5578.4}{\per\centi\metre}$, $E^6_\sigma
    = \SI{868.8}{\per\centi\metre}$ and $\xi = \SI{658.75}{\per\centi\metre}$,
    in good agreement with our data considering a different basis set and
    scalar-relativistic Hamiltonian were used.
Spin-orbit parameters for all trivalent ions can also be compared against
    recent semi-empirical estimates.\cite{lizarazo2026}
Generally, these are within seven percent of our numbers, albeit
    systematically smaller, hinting at an
    overestimation of relativistic effects with the second-order
    Douglas-Kroll-Hess Hamiltonian.\cite{wolf2002,reiher2006}

Ground state $J$-mixing in the early lanthanides Ce, Pr, Nd is limited to roughly 1\%--2\%,
    being strongest in Ce.
For the Sm and Eu compounds, mixing becomes more pronounced,
    yielding ground-state compositions
    \begin{equation*}
        \begin{split}
            &87.24\% \text{ }^{6}H_{5/2} + 10.78\% \text{ }^{6}H_{7/2} + 1.75\% \text{ }^{6}H_{9/2} + ... \text{  and} \\
            &93.41\% \text{ }^{7}F_{0}   +  4.02\% \text{ }^{7}F_{2}   + 1.75\% \text{ }^{7}F_{4} + ...,
        \end{split}
    \end{equation*}
    respectively.
$J$-mixing for the first and second excited states of Sm reduces sharply,
    amounting to 3\%, whereas for Eu, the second excited state exhibits
    a 89.5\% ${}^{7}F_{1}$ dominant configuration.
Judging from recent high-accuracy measurements on trivalent lanthanide
    ions,\cite{nielsen2025} these results indicate either an exotic electronic
    structure or that the corresponding minimal
    CASSCF wave functions and energies are not yet converged and should be
    treated at higher accuracy.

\subsection{Ln(III):\ce{Y2O3}}
\begin{figure}
    \centering
    \includegraphics[width=0.6\textwidth]{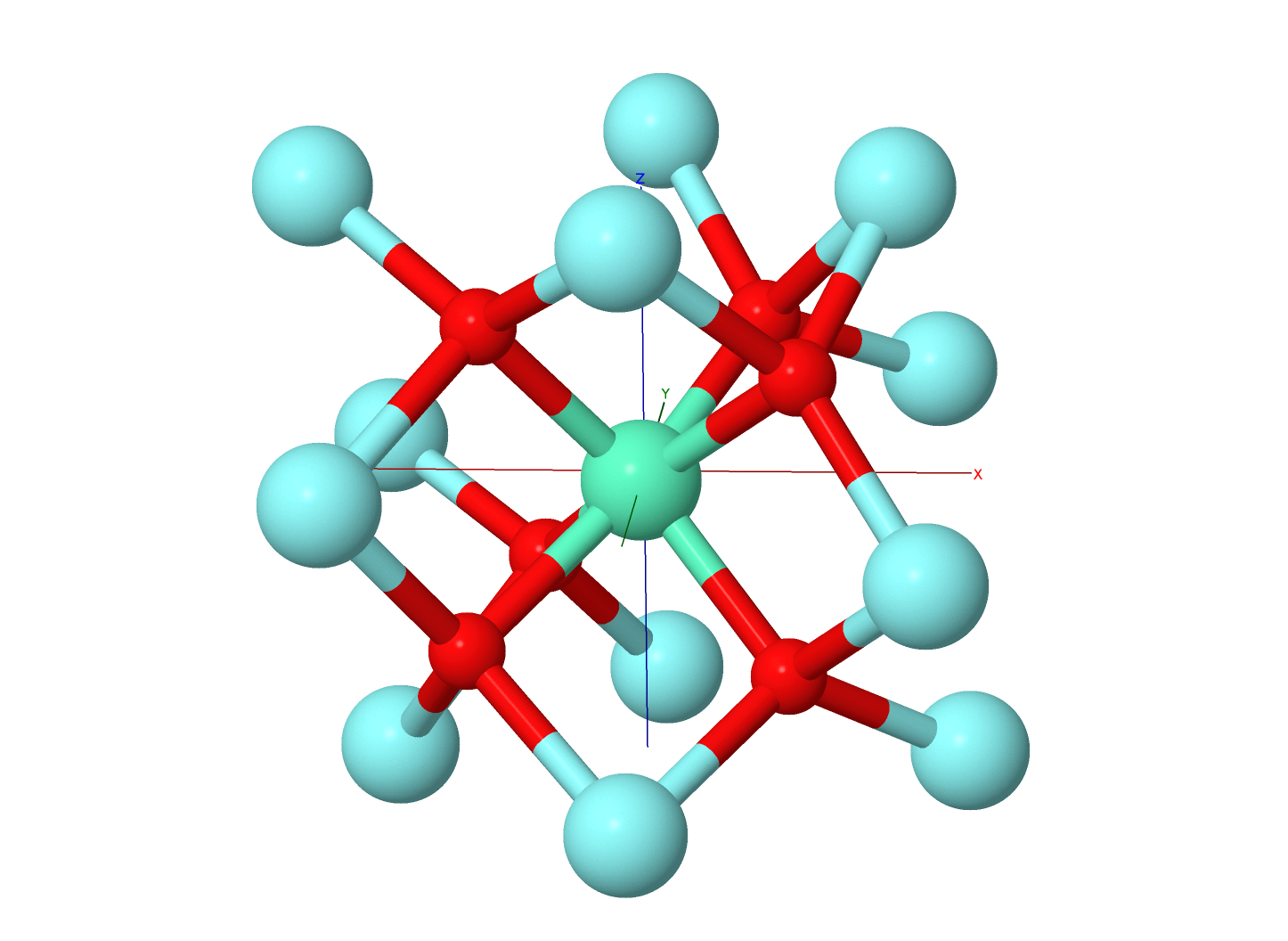}
    \caption{\ce{[Ln(III)Y12O6]^{27+}} embedded cluster model of a Ln(III):\ce{Y2O3} impurity as visualized with
        Jmol.\cite{hanson2010} \ce{Ln(III)} shown in mint turquoise, \ce{O^{2-}} in red
        and \ce{Y^{3+}} in light cyan. Point charges omitted for clarity.}
    \label{fig:impurity}
\end{figure}

\ce{Y2O3} has long been known as an excellent host material for solid-state
    lasers.\cite{morrison1982}
Within the unit cell, \ce{Y^{3+}} occurs in two distinct types with
    $C_2$ and $C_{3i}$ site symmetry, respectively.
Here, we consider lanthanide doping into the former position, see figure
    \ref{fig:impurity}, studied the most experimentally, because electric
    dipole transitions are parity-allowed.
Natural labels for this system arise under the chain $SO(3)_L \supset C_2$
    causing term splittings $F \rightarrow 3A + 4B$ (\ce{Ce^{3+}} /
    \ce{Eu^{3+}}), $H \rightarrow 5A + 6B$ (\ce{Pr^{3+}}, \ce{Sm^{3+}}) and
    $I \rightarrow 7A + 6B$ (\ce{Nd^{3+}}).

\begin{table}
    \caption{Crystal field- and spin-orbit-parameters in cm$^{-1}$ obtained for Ln(III):\ce{Y2O3} from
    least-square fits in the spin-free $\ket{S, Lt}$ and total angular momentum
    basis $\ket{J m_J}$.}
    \centering
    \renewcommand{\arraystretch}{1.1}
    \resizebox{\textwidth}{!}{%
    \begin{tabular}{L RR|RR|RR|RR|RR}
                   & \multicolumn{2}{c }{\ce{Ce^{3+}}} & \multicolumn{2}{c }{\ce{Pr^{3+}}} & \multicolumn{2}{c }{\ce{Nd^{3+}}} & \multicolumn{2}{c }{\ce{Sm^{3+}}} & \multicolumn{2}{c}{\ce{Eu^{3+}}} \\
                   & \ket{S, Lt} & \ket{J m_J} & \ket{S, Lt} & \ket{J m_J} & \ket{S, Lt} & \ket{J m_J} & \ket{S, Lt} & \ket{J m_J} & \ket{S, Lt} & \ket{J m_J} \\
                   \cmidrule{2-11}
\rowcolor{shaderow}
        E^2_\sigma & 536     & 536    & 510     & 510     & 390     & 391     & 444     & 445     & 355     & 355     \\
        E^2_{1c}   & 1287    & 1287   & 1099    & 1099    & 923     & 923     & 1018    & 1018    & 773     & 773     \\
\rowcolor{shaderow}
        E^2_{2c}   & 112     & 112    & 89      & 90      & 25      & 27      & 71      & 71      & 49      & 49      \\
        E^4_\sigma & -2673   & -2675  & -2245   & -2246   & -1731   & -1735   & -1423   & -1425   & -1214   & -1214   \\
\rowcolor{shaderow}
        E^4_{1c}   & -1993   & -1993  & -1566   & -1566   & -1194   & -1194   & -994    & -994    & -842    & -842    \\
        E^4_{2c}   & 646     & 646    & 406     & 405     & 480     & 478     & 406     & 406     & 294     & 294     \\
\rowcolor{shaderow}
        E^4_{3c}   & -3292   & -3292  & -2468   & -2468   & -1973   & -1973   & -1654   & -1654   & -1338   & -1338   \\
        E^4_{4c}   & -1930   & -1930  & -1495   & -1527   & -1226   & -1224   & -1044   & -1034   & -845    & -845    \\
\rowcolor{shaderow}
        E^6_\sigma & 664     & 658    & 587     & 579     & 570     & 569     & 483     & 483     & 406     & 406     \\
        E^6_{1c}   & -179    & -179   & -93     & -93     & 27      & 27      & 100     & 100     & 68      & 68      \\
\rowcolor{shaderow}
        E^6_{2c}   & 258     & 258    & 263     & 261     & 254     & 254     & 209     & 209     & 176     & 176     \\
        E^6_{3c}   & -399    & -399   & -250    & -250    & -218    & -218    & -150    & -150    & -112    & -112    \\
\rowcolor{shaderow}
        E^6_{4c}   & -1359   & -1359  & -1321   & -1344   & -1267   & -1268   & -1075   & -1068   & -946    & -946    \\
        E^6_{5c}   & -1054   & -1054  & -646    & -646    & -506    & -506    & -343    & -343    & -242    & -242    \\
\rowcolor{shaderow}
        E^6_{6c}   & 368     & 368    & 365     & 337     & 261     & 261     & 165     & 176     & 158     & 158     \\
        \xi        & -       & 664    & -       & 795     & -       & 944     & -       & 1244    & -       & 1417    \\
\rowcolor{shaderow}
        b          & 0       & -1     & 3       & 6       & 2       & 0       & 7       & 0       & 0       & 0       \\
        R^2        & >0.9999 & >0.9999& >0.9999 & >0.9999 & >0.9999 & >0.9999 & >0.9999 & >0.9999 & >0.9999 & >0.9999 \\
    \end{tabular}%
    \label{tab:impurity-data}
    }%
\end{table}

In point group $C_2$, all CF parameters of the type $B^k_{\lambda s}$ vanish, leaving 16 remaining parameters
    including spin-orbit coupling; a listing of these parameters is given in
    table \ref{tab:impurity-data} together with their numerical values.
While we are not aware of reference data to compare with, it is encouraging
    to see that the spin-orbit coupling parameters extracted from this system are
    almost identical to the ones extracted from the \ce{[Ln(III)(C8H8)2]-}
    complexes.
This trend is in line with the locality and atomic nature of the
    spin-orbit interaction.

Oxygen ligands are expected to cause a stronger ligand field and more $J$-mixing.
Numerically, this trend is confirmed.
Analogous to the previous section, the Ce, Pr, and Nd
    experience weaker mixing, with 97.5\% (Ce), 99.3\% (Pr) and 98.1\% (Nd) weight in the
    lowest-energy $^{2S+1}L_J$ term.
$L$-mixing in the excited states of \ce{Pr^{3+}} is still present, but amounts
    only to (1.54\%), which is roughly half of the equivalent sandwich complex in the worst case.

For Sm and Eu, we obtain the compositions
    \begin{equation*}
       \begin{split}
            &95.82\% \text{ }^{6}H_{5/2} +  2.73\% \text{ }^{6}H_{7/2} + 0.79\% \text{ }^{6}H_{9/2} + ... \text{  and} \\
            &90.73\% \text{ }^{7}F_{0}   +  7.14\% \text{ }^{7}F_{2}   + 1.72\% \text{ }^{7}F_{4} + ...,
        \end{split}
    \end{equation*}
    more in line with what would be expected from the Dieke diagram, giving
    us reason to believe that the states are well-resolved to first approximation.

\subsection{Actinide Crystal Field Theory}
Unlike the lanthanide 4f-shell, the actinide 5f-shell is not a shielded pseudo-core
    and typically exhibits covalency effects slightly weaker than the
    transition metal d-shells.
Particularly for high oxidation states, the strength of Coulomb, spin-same-orbit
    and crystal field interactions approach the same magnitude.
A robust theoretical procedure to model actinide crystal fields is therefore
    desirable.
While the above argument makes a strong case for treating the spin-orbit
    interaction variationally in intermediate coupling, it is not uncommon to analyze actinide
    complexes with bulky organic ligands using a two-step $LS$ coupling
    approach.\cite{galley2019,godsall2022,motta2023,tacconi2025}
The purpose of this section is not to assess the accuracy of such a
    methodology, but to evaluate the presented mapping
    procedure for actinides in the absence of both $S$- and $L$-mixing.

\begin{figure}[ht]
    \centering
    \includegraphics[width=0.6\textwidth]{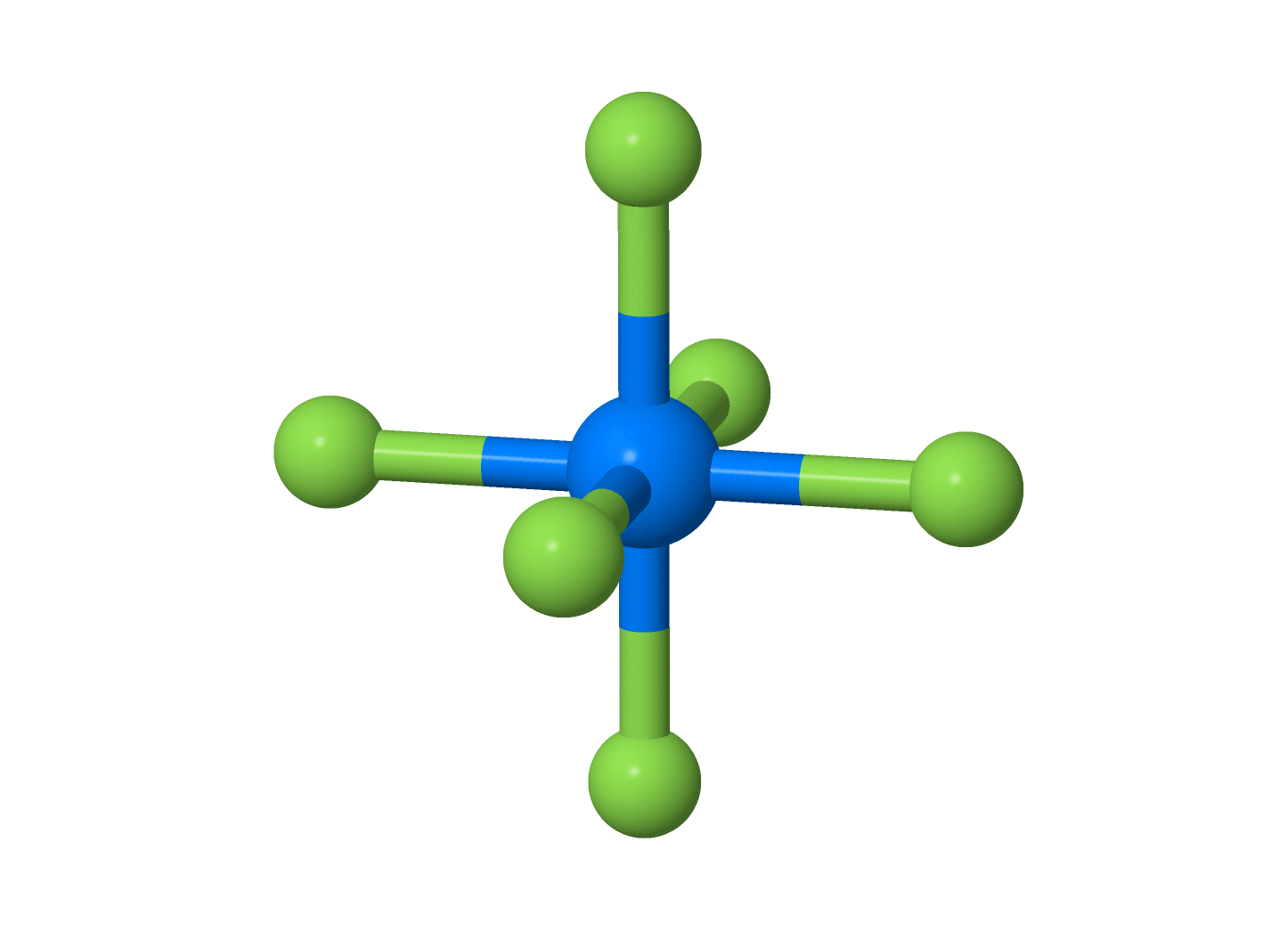}
    \caption{Octahedral \ce{NpF6} complex with a Np--F bond length of \SI{1.92}{\angstrom} as visualized with
        Jmol.\cite{hanson2010} \ce{Np^{6+}} shown in blue and \ce{F^{-}} in
        green. The coordinate axes overlap with the octahedral frame.}
    \label{fig:npf6}
\end{figure}

As an example, consider the $5f^1$ compound \ce{NpF6} with $O_h$ point group
    symmetry, shown in figure \ref{fig:npf6}.
Under the chain $SO(3)_L \supset O_h$ the atomic term splits as $F \rightarrow
    A_{2u} + T_{1u} + T_{2u}$ and the crystal field is expanded in
    the totally symmetric zonal harmonics
    $\mathfrak{H}^4_\sigma$ and $\mathfrak{H}^6_\sigma$.
Since there is only one electron with $s_i = 1/2$ and $l_i = 3$, spin and
    orbital momentum are good quantum numbers, which is a rather rare situation in the
    actinides.

\begin{table}
  \caption{Imaginary part of the numerical $L_z$ matrices in the least-square- fitted molecular
      orbital basis for the strong-field $f^1$ complex Ce:\ce{Y2O3} and \ce{NpF6} compared
      to formal atomic orbitals. Zeros are indicated by dots.}
  \centering
  \begin{tabular}{l | ccccccc}
    \multicolumn{8}{l}{formal atomic orbitals} \\
             & $\ket{3,\sigma}$ & $\ket{3,1s}$     & $\ket{3,1c}$     & $\ket{3,2s}$      & $\ket{3,2c}$      & $\ket{3,3s}$      & $\ket{3,3c}$      \\
    \midrule
    $\bra{3,\sigma}$ & $\cdot$  & $\cdot$  & $\cdot$  & $\cdot$   & $\cdot$   & $\cdot$   & $\cdot$   \\
    $\bra{3,1s}$     & $\cdot$  & $\cdot$  & 1        & $\cdot$   & $\cdot$   & $\cdot$   & $\cdot$   \\
    $\bra{3,1c}$     & $\cdot$  & $-1$     & $\cdot$  & $\cdot$   & $\cdot$   & $\cdot$   & $\cdot$   \\
    $\bra{3,2s}$     & $\cdot$  & $\cdot$  & $\cdot$  & $\cdot$   & 2         & $\cdot$   & $\cdot$   \\
    $\bra{3,2c}$     & $\cdot$  & $\cdot$  & $\cdot$  & $-2$      & $\cdot$   & $\cdot$   & $\cdot$   \\
    $\bra{3,3s}$     & $\cdot$  & $\cdot$  & $\cdot$  & $\cdot$   & $\cdot$   & $\cdot$   & 3         \\
    $\bra{3,3c}$     & $\cdot$  & $\cdot$  & $\cdot$  & $\cdot$   & $\cdot$   & $-3$      & $\cdot$   \\
    \midrule
    \multicolumn{8}{l}{} \\
    \multicolumn{8}{l}{\ce{NpF6}} \\
             & $\ket{3,\sigma}$ & $\ket{3,1s}$     & $\ket{3,1c}$     & $\ket{3,2s}$      & $\ket{3,2c}$      & $\ket{3,3s}$      & $\ket{3,3c}$      \\
    \midrule
    $\bra{3,\sigma}$ & $\cdot$  & $\cdot$   & $\cdot$   & $\cdot$   & $\cdot$   & $\cdot$   & $\cdot$   \\
    $\bra{3,1s}$     & $\cdot$  & $\cdot$   & 0.8897    & $\cdot$   & $\cdot$   & $\cdot$   & $-0.0902$ \\
    $\bra{3,1c}$     & $\cdot$  & $-0.8897$ & $\cdot$   & $\cdot$   & $\cdot$   & $-0.0902$ & $\cdot$   \\
    $\bra{3,2s}$     & $\cdot$  & $\cdot$   & $\cdot$   & $\cdot$   & 1.9203    & $\cdot$   & $\cdot$   \\
    $\bra{3,2c}$     & $\cdot$  & $\cdot$   & $\cdot$   & $-1.9203$ & $\cdot$   & $\cdot$   & $\cdot$   \\
    $\bra{3,3s}$     & $\cdot$  & $\cdot$   & 0.0902    & $\cdot$   & $\cdot$   & $\cdot$   & 2.4112    \\
    $\bra{3,3c}$     & $\cdot$  & 0.0902    & $\cdot$   & $\cdot$   & $\cdot$   & $-2.4112$ & $\cdot$   \\
    \midrule
    \multicolumn{8}{l}{} \\
    \multicolumn{8}{l}{Ce:\ce{Y2O3}} \\
             & $\ket{3,\sigma}$ & $\ket{3,1s}$     & $\ket{3,1c}$     & $\ket{3,2s}$      & $\ket{3,2c}$      & $\ket{3,3s}$      & $\ket{3,3c}$      \\
    \midrule
    $\bra{3,\sigma}$ & $\cdot$   & $-0.0011$ & $\cdot$   & 0.0043    & $\cdot$   & 0.0121    & $\cdot$   \\
    $\bra{3,1s}$     & 0.0011    & $\cdot$   & 0.9817    & $\cdot$   & 0.0019    & $\cdot$   & $-0.0042$ \\
    $\bra{3,1c}$     & $\cdot$   & $-0.9817$ & $\cdot$   & $-0.0103$ & $\cdot$   & $-0.0004$ & $\cdot$   \\
    $\bra{3,2s}$     & $-0.0043$ & $\cdot$   & 0.0103    & $\cdot$   & 1.9401    & $\cdot$   & $-0.0108$ \\
    $\bra{3,2c}$     & $\cdot$   & $-0.0019$ & $\cdot$   & $-1.9401$ & $\cdot$   & 0.0101    & $\cdot$   \\
    $\bra{3,3s}$     & $-0.0121$ & $\cdot$   & 0.0004    & $\cdot$   & $-0.0101$ & $\cdot$   & 2.9582    \\
    $\bra{3,3c}$     & $\cdot$   & 0.0042    & $\cdot$   & 0.0108    & $\cdot$   & $-2.9582$ & $\cdot$   \\
    \midrule
  \end{tabular}
  \label{tab:actinide-comparison}
\end{table}

Following the least-squares procedure established in section
    \ref{sec:uga-classification}, the orbital momentum
    matrices shown in table \ref{tab:actinide-comparison} are obtained.
Values for the Ce:\ce{Y2O3} system are provided as well to
    demonstrate the differences to a $4f$ system.
\begin{table}[ht]
    \caption{Crystal field- and spin-orbit-parameters in cm$^{-1}$ obtained for
        \ce{NpF6} from least-square fits in the spin-free $\ket{S, Lt}$ and total angular momentum
        basis $\ket{J m_J}$.}
    \centering
    \renewcommand{\arraystretch}{1.1}
    \begin{tabular}{L RR}
                   & \multicolumn{2}{c }{\ce{Np^{6+}}} \\
                   & \ket{S, Lt} & \ket{Jm_J} \\
                   \cmidrule{2-3}
\rowcolor{shaderow}
        E^4_\sigma  &  39987 &  39287 \\
        E^4_{4c}    &  33795 &  33795 \\
\rowcolor{shaderow}
        E^6_\sigma  &   5427 &   4799 \\
        E^6_{4c}    & -14358 & -14358 \\
\rowcolor{shaderow}
        \xi         & -      & 2165   \\
        b           & 0      &  553   \\
\rowcolor{shaderow}
        R^2        & >0.9999 & >0.9999
    \end{tabular}
    \label{tab:npf6-extraction}
\end{table}
Due to increased covalency, deviations from the atomic limit
    are more pronounced than for strong-field lanthanide complexes, which in itself
    is unproblematic.
However, in the effective Hamiltonian, large symmetry-forbidden matrix
    elements arise.
A fit to the corresponding numerical matrices leads to the parameters shown in
    table \ref{tab:npf6-extraction}.

The limitations of simplistic crystal field Hamiltonians for the actinides were acknowledged
    by Judd already in the 1970s.\cite{judd1977}
As a minimally invasive remedy, he suggested to introduce orbital hybridization
    variables, effectively doubling the number of crystal field parameters.
Our calculations indeed confirm two additional parameters in place of
    $E^4_{4c}$ and $E^6_{4c}$; yet, without re-deriving the model including the
    proper pre-factors these numbers cannot be interpreted.
In the $\ket{J m_J}$ basis, orbital hybridization causes a large
    barycenter correction, indicating the inappropriateness of the
    present crystal field model.
Moreover, strong $J$-mixing is observed across the spectrum,
    with a ground-state composition $87.55\% {}^2F_{5/2} / 12.45\% {}^2F_{7/2}$, questioning
    the validity of extracting CF parameters from a single $J$-manifold in actinides.
How to account for these covalency effects in general actinide $f^n$
    configurations will be examined in future work.

\section{Conclusions}
We presented a protocol to extract CF parameters from CASSCF calculations that is applicable to
    the lanthanide $J$-mixing limit and based on a synthesis of real tensorial sets in
    $\mathbb{R}^3$ with the unitary group approach.
Historically, the real tensorial formalism was motivated by what became
    known as the angular overlap model, a perturbative extension of traditional
    crystal field theory to covalent bonding.\cite{schaeffer1970,schaeffer1971,schaeffer1973}
The angular overlap model is particularly suited to inform experimental ligand
    design of transition metal and actinide complexes, where covalency effects
    cannot be neglected.
The tools developed as part of this work allow to parametrize angular overlap
    models based on CASSCF calculations for the transition metal $d$- and
    lanthanide $f$-shells in arbitrary geometries and this deserves further study.
Another natural extension concerns the model treatment of $4f/5d$ interactions
    to account for the dominant corrections to the minimal $f^n$-only
    treatment.\cite{rajnak1963}
In this context it will also become necessary to reconsider how chemical
    insights may be extracted from an increasingly large parameter set.

\begin{acknowledgement}
    A.A.S. acknowledges funding by the Deutsche Forschungsgemeinschaft (German
    Research Foundation, DFG) under project number 578382806 and by the Swiss
    National Science Foundation (SNSF) under project number 200021\_219616.
\end{acknowledgement}

{ \footnotesize \singlespacing

\providecommand{\latin}[1]{#1}
\makeatletter
\providecommand{\doi}
  {\begingroup\let\do\@makeother\dospecials
  \catcode`\{=1 \catcode`\}=2 \doi@aux}
\providecommand{\doi@aux}[1]{\endgroup\texttt{#1}}
\makeatother
\providecommand*\mcitethebibliography{\thebibliography}
\csname @ifundefined\endcsname{endmcitethebibliography}
  {\let\endmcitethebibliography\endthebibliography}{}

}

\end{document}